\documentclass[twocolumn, secnumarabic,amssymb, nobibnotes, aps, prd, dbfloatfix]{revtex4-2}

\usepackage{ragged2e}
\usepackage[justification=justified,singlelinecheck=false]{caption}
\DeclareCaptionJustification{justified}{\justifying}
\usepackage{siunitx}
\usepackage{amstext}
\usepackage{amssymb}
\usepackage{amsmath}
\usepackage{float}

\usepackage[version=4]{mhchem}
\usepackage{siunitx}
\DeclareSIUnit{\wtpercent}{wt.\%}
\DeclareSIUnit{\atpercent}{at.\%}
\DeclareSIUnit{\sccm}{sccm}
\usepackage{leftindex}
\usepackage{mathtools}

\usepackage{nicefrac,xfrac}
\usepackage{booktabs}
\usepackage{multirow}

\usepackage{color,soul}

\usepackage[labelfont=bf]{caption}
\usepackage{graphicx}
\usepackage{epstopdf}
\DeclareGraphicsExtensions{.eps,.pdf}

\usepackage[colorlinks, allcolors = blue]{hyperref}

\usepackage{leftindex}
\expandafter\let\csname equation*\endcsname\relax
\expandafter\let\csname endequation*\endcsname\relax

\usepackage{amstext}
\usepackage{amssymb}
\usepackage{amsmath}

\usepackage[version=4]{mhchem}
\begin{document}

\title[]{On the crystalline environment of luminescent \mbox{Tb$^{3+}$} ions embedded in indium tin oxide thin films: a DFT and Crystal field analysis assessment}

\author{E. Serquen$^{1}$}\email[Contact author: ]{esserquen@pucp.edu.pe}
\author{K. Lizárraga$^{1,2}$}
\author{L. A. Enrique$^{1}$, F. Bravo$^{1}$, S. Mishra$^{1}$, P. LLontop$^{3}$, P. Venezuela$^{2}$, L. R. Tessler$^4$}
\author{J. A. Guerra$^1$}\email[Contact author: ]{guerra.jorgea@pucp.edu.pe}

\affiliation{$^1$Departamento de Ciencias, Secci\'on F\'isica, Pontificia Universidad Cat\'olica del Per\'u, Av. Universitaria 1801, 15088, Lima, Per\'u}
\affiliation{$^2$Instituto de F\'isica, Universidade Federal Fluminense, 24210-346, Niterói, RJ, Brazil}
\affiliation{$^3$MESA+ Institute for Nanotechnology, University of Twente, Enschede 7500 AE, The Netherlands.}
\affiliation{$^4$Instituto de F\'isica Gleb Wataghin, Universidade Estadual de Campinas, 13083-859, Campinas, SP, Brazil}
\date{\today}
\begin{abstract}
We assess the local symmetry and crystal environment of trivalent terbium ions embedded in an indium tin oxide (ITO) matrix with bixbyite structure. The \mbox{Tb$^{3+}$} ions tend to substitute \mbox{In$^{3+}$} ions in two different cationic sites ($b$ and $d$). Density Functional Theory (DFT) calculations suggest that the \mbox{Tb$^{3+}$} ions are mainly located at $C_2$ symmetry sites relaxing selection rules and enabling electric dipole transitions, with the $^5\text{D}_4\rightarrow\leftindex^7{\text{F}}_2$ transition being the most intense, providing a red color to the light emission. Photoluminescence emission spectra under UV excitation at \qty{83}{\kelvin} revealed 30 intra-4$f$ transitions, which were assigned to the $\leftindex^7{\text{F}}_J$ ground multiplet of the \mbox{Tb$^{3+}$} ion. Crystal-field analysis shows a strong alignment between calculated and observed energy levels, yielding a standard deviation of $\sigma=\qty{15.1}{\centi\per\metre}$. We believe these results can help to understand the activation mechanisms of \mbox{Tb$^{3+}$} luminescent centers in transparent conductive oxides, as well as the potential to modulate \mbox{Tb$^{3+}$} emission color through its crystalline environment.
\end{abstract}

\maketitle
\section{Introduction}
Trivalent rare earth (RE) ions have the capability to emit light when incorporated in some transparent host materials covering a wide range of the visible spectrum. Their distinctive spectroscopic properties primarily arise from electronic transitions within their partially filled $4f^n$ orbitals ($n$=0-14). Intra-$4f$ electric dipole (ED) transitions are forbidden by Laporte selection rules and only magnetic dipole (MD) transitions are allowed in the free ion case. However, when RE ions are embedded into a matrix occupying a non-centrosymmetric site in the host lattice, the interactions with the crystal field (CF) mix electronic states having opposite parity. The free ion 4$f$ energy levels split under the influence of the local electric field produced by the crystalline environment, with the point symmetry around the ion being the only factor that determines how much the degeneracy of the free-ion state is removed \cite{Liu2005, Walsh2006}.

Among RE ions, terbium (\mbox{Tb$^{3+}$}) has been widely used as a dopant of wide and ultra-wide bandgap semiconductor host materials, such as SiC, Ga$_2$O$_3$, AlN, ZnO, a-SiN:H, and GaN \cite{guerra2013, WANG2021,Guerra_2016, GUO2023127361, Bosco18, braun2022closing}. The energy transfer from the host to the \mbox{Tb$^{3+}$} ions enables them to act as functional luminescent centers, with the \mbox{$^5\text{D}_4\rightarrow\leftindex^7{\text{F}}_5$} transition in the green regions around \qty{545}{\nm} being typically the most intense. This makes \mbox{Tb$^{3+}$} ions ideal to be used as green phosphorus for photon downshift layers, as well as down- and up-conversion layers, biomarkers, fluorescent lamps, optical fiber temperature sensors, and LEDs \cite{DEVI2020, petoral2009, tahiri2020, Zhu2019, Guo2024}. Nevertheless, the emission color of the \mbox{Tb$^{3+}$} ions can be influenced by the chemical environment and the excitation process. For instance, it has been previously reported the capability to tune the color emission of \mbox{Tb$^{3+}$} ions  by reducing the Tb concentration in the host matrix, leading to an increase in the  blue emission, related to  $^5\text{D}_3\rightarrow\leftindex^7{\text{F}}_{4,5,6}$ transitions, while all other transitions intensities are quenched \cite{de2003tb3+, DOSSANTOS2021118430, Benz2013}. 

In the last decade, Transparent Conductive Oxides (TCOs) have emerged as promising host materials for RE ions since they combine high optical transmittance in the UV-Vis range and electrical conductivity, making them key components in many optoelectronic devices \cite{MaL2015}. The current trend of incorporating RE ions into TCOs aims to add luminescent properties to these materials. Thus, different host of TCOs based on In$_2$O$_3$, SnO$_2$, CdO, Ga$_2$O$_3$, and ZnO have been doped with several RE ions \cite{Pereira2006, Sakthivel2019, COLAK2018, GILROSTRA2020, CHOI2008290}. One of the most investigated TCOs host materials is indium tin oxide (ITO; SnO$_2$ doped In$_2$O$_3$) \cite{CHOI2008290, KIM2009, Sunde2014}.
The impact of \mbox{Tb$^{3+}$} incorporation on the electrical, optical, and light emission properties of ITO thin films has been recently investigated \cite{Llontop_2022}. The latter study reported an unusually intense red-light emission from the \mbox{Tb$^{3+}$} ions, associated to \mbox{$^5\text{D}_4\rightarrow\leftindex^7{\text{F}}_{2}$} transition. This transition can be resolved into five peaks due to Stark splitting, highlighting the crystal field effect on the \mbox{Tb$^{3+}$} ion in a low symmetry site.

\begin{figure*}[t!] 
    \centering
    \includegraphics[width=0.45\textwidth]{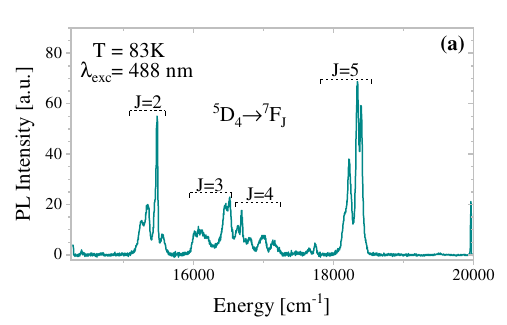}
    \includegraphics[width=0.45\textwidth]{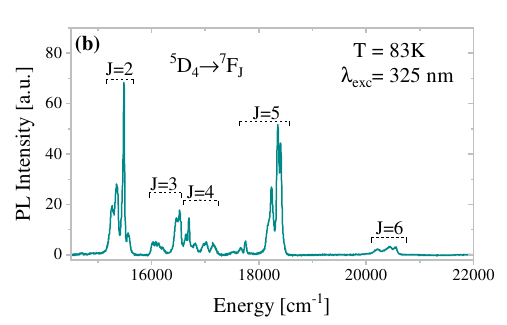}
    \caption{Photoluminescence spectra at \qty{83}{\kelvin} under excitation wavelength of \qty{488}{\nm} (a), and \qty{325}{nm} (b). $^5\text{D}_4$ related electronic transitions are denoted.}
    \label{Pl_spectra}     
\end{figure*}

In this work, we present experimental results validated by a theoretical analysis based on density functional theory (DFT) calculations and crystal field (CF) analysis to identify the local symmetry of the lattice site occupied by the \mbox{Tb$^{3+}$} ions in the ITO lattice with a bixbyite structure. Low temperature photoluminescence under UV excitation presented a multiple peak structure. The individual peaks were associated to transitions within the $^7\text{F}_{JM}$ energy levels corresponding to $C_2$ symmetry sites. Results show the remarkable relation between the local symmetry and \mbox{Tb$^{3+}$} emission color. These results may be applied to host with similar symmetries allowing the tunability of the \mbox{Tb$^{3+}$} light emission color.

\section{Experimental details}
The experimental procedures, including the synthesis method and characterization techniques can be found elsewhere  \cite{Llontop_2022}. For the sake of brevity, only a summary of the key aspects is provided here. ITO:Tb thin films were synthesized by magnetron sputtering technique using ITO (\qty{90}{\wtpercent} In$_2$O$_3$ \& \qty{10}{\wtpercent} SnO$_2$) and Tb targets. Photoluminescence activation was achieved after annealing at \qty{650}{\degreeCelsius} in air.

Photoluminescence (PL) spectra were recorded in a Renishaw inVia micro-Raman spectrometer with HeCd (\qty{325}{\nm}) and Ar (\qty{448}{\nm}) laser excitation. Low temperature PL measurements were carried out in a Linkam THMS600 cooling stage with a liquid nitrogen inlet.

\section{Results and discussion}
\subsection{\mbox{Tb$^{3+}$} emission spectrum}
PL spectra of the sample containing \qty{1.5}{\atpercent} Tb at \qty{83}{\kelvin} under \qty{488}{\nm} and \qty{325}{nm} excitation are shown in \mbox{Figure \ref{Pl_spectra}}. The observed \mbox{Tb$^{3+}$} related emission corresponds to the  \mbox{$^5\text{D}_4\rightarrow\leftindex^7{\text{F}}_J$} transitions. \mbox{Figure \ref{Pl_spectra}\textcolor{blue}{(a)}} depicts the PL spectrum under sub-bandgap excitation with a \qty{488}{\nm} laser. This excitation is quasi-resonant with \mbox{$^5\text{D}_4\rightarrow\leftindex^7{\text{F}}_6$} transition. On the other hand, Figure \ref{Pl_spectra}\textcolor{blue}{(b)} depicts the PL spectrum under a \qty{325}{\nm} wavelength excitation, which corresponds to sub-bandgap excitation involving band tail states  ($E_\text{g} = \qty{3.82}{\eV}$, $E_\text{U}=\qty{0.22}{\eV}$) \cite{Llontop_2022}. We did not observe any transitions associated with the $^5\text{D}_3$ initial states. Since, the energy difference between the $^5\text{D}_3$ and $^5\text{D}_4$ level is approximately $\qty{5600}{\cm^{-1}}$ \cite{Li2007}, and the phonon threshold in In$_2$O$_3$ is around $\qty{300}{\cm^{-1}}$ \cite{Irmscher2014}, excited ions at $^5\text{D}_3$ state could decay to the $^5\text{D}_4$ state mediated by multiphonon relaxation. However, at these Tb concentrations, a cross-relaxation process may also take place by the interaction between two neighboring \mbox{Tb$^{3+}$} ions, leading to the quenching of $^5\text{D}_3$ related light emission as a result of the energy transfer from \mbox{$^5\text{D}_3\rightarrow\leftindex^5{\text{D}}_4$} to \mbox{$^7{\text{F}}_J$} manifold \cite{DOSSANTOS2021118430,Benz2013}.

To determine the fractional contribution of a specified measurable transition ( \mbox{$^5\text{D}_J\rightarrow\leftindex^7{\text{F}}_{J'}$}) to the total radiative transition probability from the excited state, the relative intensities for each transition ($\beta_{JJ'}$) were calculated according to:
\begin{equation}
    \beta_{JJ'} =\dfrac{I_{JJ'}}{\sum I_{JJ'}}
\end{equation}
Here $I_{JJ'}$ is the integrated emission intensity of a specific  $J\rightarrow J'$ transition. The selection rules on the total angular quantum number $J$ can help to identify the transition mechanism, i.e. electric dipole (ED) or magnetic dipole (MD) mediated. The $J=0\rightarrow J'=0$ transition is forbidden by ED and MD. MD transitions are subjected to the selection rules $\Delta S= 0$, $\Delta L= 0$, $\Delta J= 0, \pm 1$. Whereas, for ED transitions the selection rules $\Delta S= 0$, $\Delta L\leq 6$, and $\Delta J\leq 6$ apply, which can be further restricted to $\Delta J = 2, 4, 6$ when either $J=0$ or $J'=0$ \cite{GORLLERWALRAND1998}. Table \ref{bratios} summarizes the $\beta_{JJ'}$ calculated from the spectrum measured under UV excitation, and the related transition mechanism.
 
\begin{table}[h!]      
      \caption{Fractional contribution $\beta_{JJ'}$ calculated from the spectrum under \qty{325}{\nm} excitation wavelength.}
    \begin{ruledtabular}
    \begin{tabular}{llcll}
     Transition & \hspace{0.25cm} & $\beta_{JJ'}$ [$\%$] & \hspace{0.25cm} & Type \\
     \hline
     $^5\text{D}_4\rightarrow\leftindex^7{\text{F}}_6$&&2.95&&  ED \\
     \mbox{$^5\text{D}_4\rightarrow\leftindex^7{\text{F}}_5$}&&28.53&& ED + MD  \\
     $^5\text{D}_4\rightarrow\leftindex^7{\text{F}}_4$&&10.67&& ED + MD \\
     $^5\text{D}_4\rightarrow\leftindex^7{\text{F}}_3$&&15.03&&  ED + MD\\
     $^5\text{D}_4\rightarrow\leftindex^7{\text{F}}_2$&&42.82&&  ED\\
     
    \end{tabular}
    \end{ruledtabular}
    \label{bratios}
\end{table}

Based on the selection rules, the \mbox{$^5\text{D}_4\rightarrow\leftindex^7{\text{F}}_5$} transition has a ED+MD mixed contribution, being typically the most prominent emission line. Notwithstanding, here the \mbox{$^5\text{D}_4\rightarrow\leftindex^7{\text{F}}_2$} ED transition exhibits the highest intensity in our PL spectra. Such behavior might be caused by the high sensitivity of this transition to the inhomogeneities in the dielectric surrounding of the \mbox{Tb$^{3+}$} ions. This phenomenon has been previously predicted by Richardson et al. \cite{Richardson1981} by comparing the intensity ratios of the \mbox{Tb$^{3+}$} electronic transitions in two different aqueous solutions. There, they suggested that transitions with an \textit{even} $\Delta J$ are more sensitive to the ligand environment. Electronic transitions with this behavior are called hypersensitive and have been experimentally observed to follow the selection rule \mbox{$\Delta J = \pm 2$} in several aqueous solutions \cite{Jorgensen1964}. Judd \cite{Judd1966} proposed that this effect is possible just for some symmetry groups: $C_s$, $C_1$, $C_2$, $C_3$, $C_4$, $C_6$, $C_{2v}$, $C_{3v}$, $C_{4v}$ and $C_{6v}$, although some exceptions have been found \cite{BINNEMANS20151}. Therefore, the \mbox{$^5\text{D}_4\rightarrow\leftindex^7{\text{F}}_2$} transition could be considered hypersensitive and its intensity would strongly depend on the crystal field surrounding the \mbox{Tb$^{3+}$} ions. 

\subsection{Density functional theory calculations}
We used DFT calculations to determine the preferred Tb sites in the lattice. Simulations were carried out using the Vienna Ab initio Simulation Package (VASP) \cite{kresse,kresse2}. To describe the exchange correlation density functional, we used the Perdew-Bruke-Enzerhof (PBE) generalized gradient approximation (GGA). All the calculations were performed with spin polarization, a cutoff energy of \qty{400}{\eV}, energy threshold of \qty{e-6}{\eV}, and full relaxation of atomic positions with a force threshold of \qty{0.01}{\eV/\angstrom}. A $k$-points grid of \mbox{$3\times3\times3$} following the Monkhorst Pack scheme was used to sample the Brillouin zone, and  a smearing of \qty{0.02}{\eV} was applied.

\begin{figure}[h!]
    \centering
    \includegraphics[width=0.8\columnwidth]{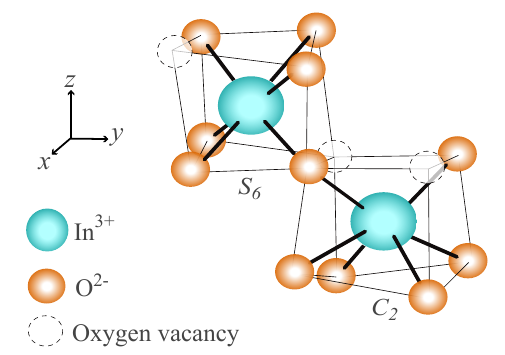} 
    \caption{Schematic presentation of the $C_2$ and $S_6$ symmetry sites in a cubic bixbyite. Used coordinate axis for the crystal field analysis are shown. The rotational axis is aligned along the $z$-direction.}
    \label{figSymmetry}
\end{figure}

The starting point of our calculations was the study of In$_2$O$_3$ bixbyite structure carried by Marezio \cite{Marezio}, from which the unit cell was obtained. The cubic bixbyite structure has two non-equivalent cationic positions corresponding to $d$ and $b$ sites in the Wickoff notation, each coordinated to six oxygen atoms. The $d$ sites are low-symmetry positions where the cation is locateded at the center of a distorted cube  with two oxygen vacancies along one of its face diagonals, corresponding to a $C_2$ symmetry with the rotation axis aligned with the $z$-axis. In contrast, the $b$ sites are higher-symmetry positions, where the cation sits at the center of a cube with two oxygen vacancies along one of its body diagonals, leading to a trigonally compressed octahedra \cite{NADAUD1998}. The latter site corresponding to an $S_6$ symmetry with the rotation axis parallel to $z$-axis, as illustrated in Figure \ref{figSymmetry}.

\begin{figure*}[tb!]
   \centering
     \includegraphics[width=0.98\textwidth]{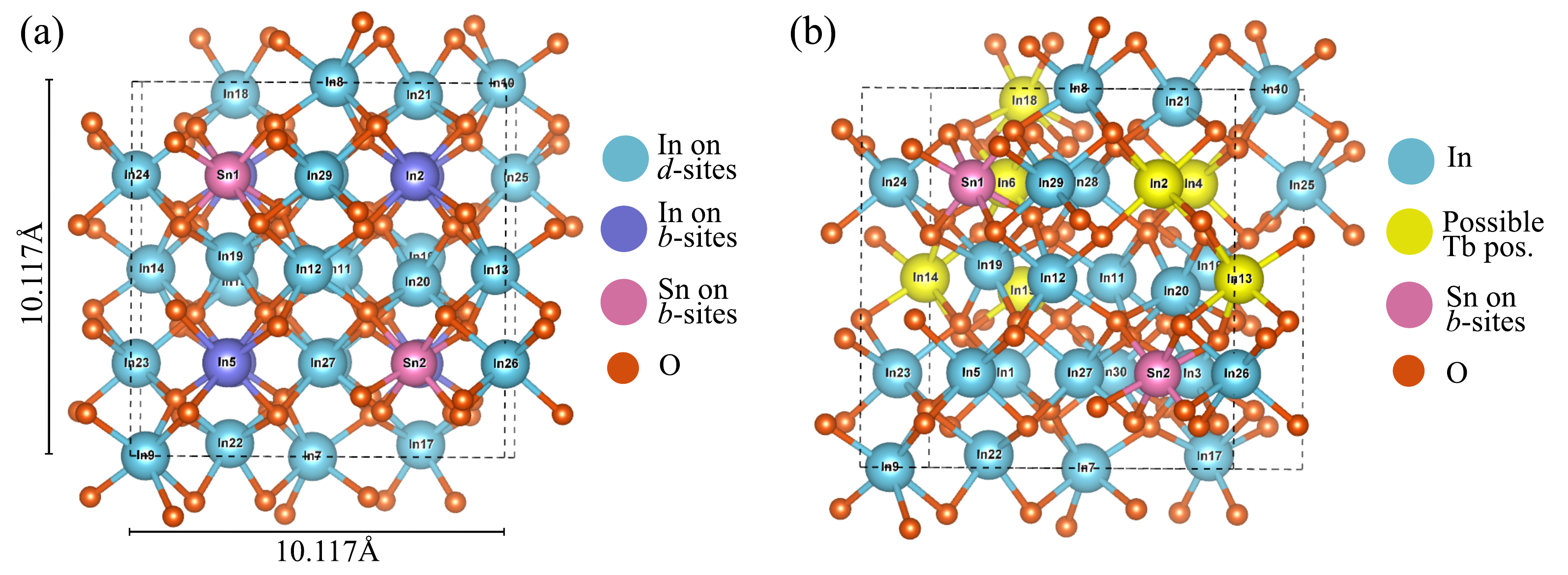}
    \caption{ITO bixbyite structure with 6$\%$ of Sn located on $b$ sites, In$_{30}$Sn$_2$O$_{48}$. Indium sites are marked with sky-blue and purple colors to point distinction of $b$ and $d$ sites (a). Possible positions of Tb within the ITO matrix are colored as yellow (b).}
    \label{fig:struct}
\end{figure*}

When In$_2$O$_3$ is doped with SnO$_2$ to form ITO, the Sn$^{4+}$ ion substitutes for the In$^{3+}$ ions, preferentially occupying $b$ sites. Then, Sn$^{4+}$ contributes with an extra electron, increasing the carrier density while preserving the cubic structure. Tripathi et al. \cite{madhvendra} studied the preferable positions of Sn impurities within the In$_2$O$_3$ structure. They conclude that the $b$ position is preferable when compared to $d$ sites. \mbox{Figure \ref{fig:struct}\textcolor{blue}{(a)}} shows the ITO unit cell with 6$\%$ of Sn in the $b$ sites. The atomic coordinates of this system can be found in the supplementary information. 

When RE elements as Er, Nd, Eu or Tb are added, the ITO bixbyite structure remains stable as long as the solubility limit is not exceeded \cite{CHOI2008290, Sunde2014}. Moreover, a theoretical analysis using first-principles methods, performed by Stanek et al. \cite{Stanek2007}, showed that trivalent RE dopants in oxides with the bixbyite structure tend to localize at $d$ sites when the solute cation is smaller than the host lattice cation, as this is energetically favorable. However, when the solute cation is larger than the host lattice cation, the $b$ sites may also be preferred. In case of Tb-doped ITO, Tb ions can substitute either In or Sn ions in the $b$ or $d$ sites, inducing stress in the lattice due to the fact that the ionic radius of the six-coordinate \mbox{Tb$^{3+}$} ion (r=\qty{0.92}{\angstrom}) \cite{Templeton, Gupta2016}  is larger than the radius of both In$^{3+}$(r=\qty{0.80}{\angstrom}) and Sn$^{4+}$(r =\qty{0.72}{\angstrom}) \cite{MENG2022}.
Indded, the as-grown Tb-doped ITO layer exhibited an amorphous structure due to the active cooling of the sample holder during growth. After annealing at \qty{550}{\degreeCelsius}, it crystallizes into the cubic bixbyite-type ($Ia\bar{3}$) structure \cite{Llontop_2022}.

Here we performed the simulation of a unit cell with 80 atoms with $8b$ sites and $24d$ sites. The 3\% of Tb incorporation was done by substituing In atoms only, keeping the 6\% of Sn in ITO. This results in 30 possible positions for Tb. Nevertheless, many of these configurations have equivalent energy states. To differentiate them, we calculate the distance between Tb and both Sn atoms. This approach left 7 configurations with different in energies, three with Tb at $b$ sites (Tb$_{\textrm{In}2}$, Tb$_{\textrm{In}4}$, and Tb$_{\textrm{In}6}$), and four with Tb at $d$ sites (Tb$_{\textrm{In}13}$, Tb$_{\textrm{In}14}$, Tb$_{\textrm{In}15}$, and Tb$_{\textrm{In}16}$). These can be seen in \mbox{Figure \ref{fig:struct}\textcolor{blue}{(b)}} where the possible Tb positions are highlighted in yellow. 
The relative energies of the seven configurations can be found in Table \ref{tab:relative.energy}, where the lowest energy corresponds to the Tb$_{\textrm{In}14}$ system. This indicate that Tb atoms prefer $d$ sites over $b$ sites. In addition, we have calculated the relative probability of a given state in thermodynamic equilibrium at certain temperature given by the Boltzmann factor, $p_i$, i.e.,
\begin{equation}
    p_i=\exp\left(\nicefrac{E_{i}}{k_\beta T}\right)
\end{equation}

The relative probabilities for each configuration was calculated for \qty{300}{\kelvin} and \qty{923}{\kelvin}, representing room and annealing temperatures, respectively. The results show that at room temperature, the Tb atoms prefer $d$ sites. Although at \qty{923}{\kelvin} there is an enhancement of the probability for both sites, $d$ sites are still the most probable ones.

\begin{table}[h!]
    \centering
        \caption{Relative energies ($E_{i}$) and Boltzmann probability factor ($p_{i}$) calculated at \qty{300}{\kelvin} and \qty{923}{\kelvin}.}
   \begin{ruledtabular}
     \begin{tabular}{lccccc}
    \multirow{2}{*}{System} & $E_{i}$ & $p_{i}$ & $p_{i}$& \multirow{2}{*}{Site}& \multirow{2}{*}{Symmetry}\\
    & [$\unit{\milli\eV}$] & at \qty{300}{\kelvin} &  at \qty{923}{\kelvin} & &\\
    \hline
    Tb$_{\textrm{In}2}$& 30 & 0.31  & 0.69 & $b$ & $S_6$\\
    Tb$_{\textrm{In}4}$& 59 & 0.10  & 0.48 & $b$ & $S_6$\\
    Tb$_{\textrm{In}6}$& 60 & 0.10  & 0.47 & $b$ &$S_6$\\
    Tb$_{\textrm{In}13}$& 17 & 0.52  & 0.81 & $d$ & $C_2$\\
    Tb$_{\textrm{In}14}$& 0 & 1.00  & 1.00 & $d$ & $C_2$\\
    Tb$_{\textrm{In}15}$& 5 & 0.82  & 0.94 & $d$ & $C_2$\\
    Tb$_{\textrm{In}18}$& 7 & 0.76  & 0.92 & $d$ & $C_2$\\
     \end{tabular}
     \end{ruledtabular}
     \label{tab:relative.energy}
 \end{table}
 
\subsection{Crystal field parametrization of the energy levels of \mbox{Tb$^{3+}$} ions in $C_2$ symmetry sites}
The complete Crystal Field (CF) energy levels scheme of the \mbox{Tb$^{3+}$} ion consists in a total of 3003 Stark levels \cite{ANTICFIDANCEV2004}. However, here we have only considered the first 49 lowest $^7\text{F}_{JM}$ set of levels. Since the CF operators only mix levels with the same multiplicity, in this case $^7\text{F}_J$. Furthermore, the substantial energy gap between the $^5\text{D}_4$ excited state and the ground $^7\text{F}_J$ levels makes $J$-mixing effects on the excited states negligible \cite{BOSCO2019100028, ANTICFIDANCEV2004}, thus the latter approximation is valid.

\begin{figure}[hbt!]
    \centering
    \includegraphics[width=\columnwidth]{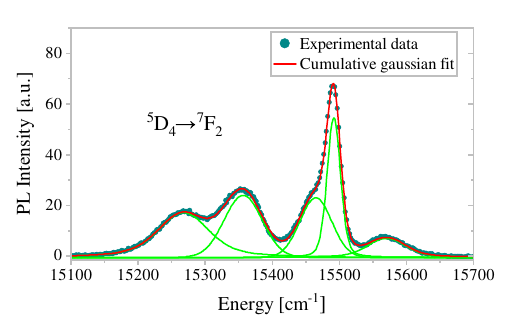}
    \caption{High resolution spectrum of the $^5\text{D}_4\rightarrow\leftindex^7{\text{F}}_2$ transition at \qty{83}{\kelvin}. The corresponding five peaks have been deconvoluted as explained in the text.}
    \label{5D4-7F2}
\end{figure}

The luminescence spectrum at \qty{83}{\kelvin} under UV excitation was carefully fitted using the sum of 30 gaussian curves. The most energetic transition was observed around \qty{20550}{\per\cm} corresponding to the \mbox{$^5\text{D}_4\rightarrow\leftindex^7{\text{F}}_6$} transition. Consistency criteria were imposed to improve the fitting accuracy. For each $2J+1$ level, the full width at half maximum (FWHM) has to decrease as the transition energy increases \cite{Taboada1994, BOSCO2019100028}, as shown for the fitted parameters of the \mbox{$^5\text{D}_4\rightarrow\leftindex^7{\text{F}}_{2}$} transition listed in \mbox{Table \ref{gauss_parameters}}. Furthermore, this transition was resolved in a maximum of five peaks, as evidence in Figure \ref{5D4-7F2}, due to Stark splitting of the $^7{\text{F}}_2$ multiplet, indicating that the \mbox{Tb$^{3+}$} ions are located at low-symmetry sites \cite{deBette2012}. 

All energy levels were determined relative to the ground state at $^7\text{F}_6$, assuming that all visible transitions originate from a single $^5\text{D}_4$ level at \qty{20550}{\per\cm} ($\sim\qty{2.55}{\eV}$). This is confirmed by the PL measurements under 488nm $(\sim\qty{2.54}{\eV})$ excitation, which are quasi-resonant with the \mbox{$^5\text{D}_4\rightarrow\leftindex^7{\text{F}}_{6}$} transition. In addition, the number of peaks in which the \mbox{$^5\text{D}_4\rightarrow\leftindex^7{\text{F}}_{2}$} can be resolved suggest that the initial \mbox{$^5\text{D}_4$} state is not splited . 

\begin{table}[hbt!]
    \centering
     \caption{Best fitted parameters of the Gaussian curves used to fit the \mbox{$^5\text{D}_4\rightarrow\leftindex^7{\text{F}}_2$} transition at \qty{83}{\kelvin}.}
    \begin{ruledtabular}
    \begin{tabular}{lccc}
  Transition & center & FWHM & Integrate intensity\\
   & [$\unit{\per\centi\metre}$]& [$\unit{\per\centi\metre}$] & [a.u.]\\
     \hline
   \mbox{$^5\text{D}_4\rightarrow\leftindex^7{\text{F}}_2$}  &15260&	83&	146374\\
    &15355&	74&	200934\\
    &15469&	54&	163258\\
    &15494&	23&	121184\\
    &15570 &66&52505\\
    \end{tabular}
    \end{ruledtabular}
    \label{gauss_parameters}
\end{table}

The experimental energy levels of \mbox{Tb$^{3+}$} were fitted using a parametric Hamiltonian for the $4f^8$ configuration, which can be written as a free-ion (fi) part and crystal field (CF) parts  \cite{Carnall89}: 
\begin{equation}
    H= H_{\text{fi}} +H_{\text{CF}}
\end{equation}  
Here the free-ion Hamiltonian $H_{\text{fi}}$ includes all spherically symmetric interactions, and is expressed as:   
\begin{equation}
\resizebox{\columnwidth}{!}{$\begin{split}
    H_{\text{fi}} = E_{\text{avg}} & + \sum_{k=2,4,6}F^k\mathbf{f_k} + \zeta_{4f}\mathbf{A_{SO}} \\ 
    & + \alpha L(L+1) + \beta G(G_2) + \gamma G(R_7)  \\
    & + \sum_{i=2,3,4,6,7,8}T^i t_i + \sum_{k=0,2,4}M^km_k + \sum_{k=2,4,6}P^kp_k.\end{split}$}
\end{equation}
Here the first term, $E_{\text{avg}}$, represents the barycenter energy of the $f^8$ configuration. The second term describes the Coulomb interactions between the $4f^8$ electrons, splitting the $f^8$ configuration into $SL$ terms. $F^k$ are the Slater parameters, and $\mathbf{f_k}$ represents the angular part of the electrostatic interaction operator. The third term accounts for spin-orbit (SO) coupling, involving the SO constant $\zeta_{4f}$ and the angular part $\mathbf{A_{SO}}$ of the interaction operator. The last six terms involve up to 15 correction and correlations factors. Where $\alpha$, $\beta$, $\gamma$ are the two-particle conﬁguration interaction parameters. Here $L$ is the total orbital angular momentum, $G(G_2)$ and $G(R_7)$ are Casimir operators for groups $G_2$ and $R_7$. The $T^i$ parameters are associated with the three-particle operators $t_i$. $M^k$ correspond to Marvin integrals with $m_k$ as their associated operators. $P^k$ are electrostatic correlated SO interaction parameters with $p_k$ being the associated operators.  Although these latter terms provide only minor corrections and do not induce energy level splitting, omitting them would compromise the accuracy of the fit. Further details are available in \cite{Carnall89, REID2016, brik2019theoretical, newman2000crystal}.

Following Wybourne formalism, the crystal field Hamiltonian $H_{\text{CF}}$ can be written as \cite{wybourne1965spectroscopic}:
\begin{equation}
\begin{split}
 H_{CF}=\sum_{k}\sum^k_{q=-k}\Big[ B^k_q \left[C^k_q + (-1)^qC^k_{-q}\right] + \\ iS^k_q \left[C^k_q +(-1)^qC^k_{-q}\right]\Big].
\end{split}
\end{equation}

where $C_q^k$ are the spherical tensor operators for the $4f^n$ configuration. The $B_q^k$ and $S_q^k$ are the real and imaginary parts of the CF parameters with $k=0,2,4,6$ and $q\neq 0$. The CF parameters fitting was performed using the $f$-shell empirical programs provided by M. F.Reid \cite{MReid84}.

Based on the observed \mbox{$^5\text{D}_4\rightarrow\leftindex^7{\text{F}}_{2}$} and DFT calculations, we assume that the emission of the Tb-doped ITO layer is due to \mbox{Tb$^{3+}$} ions mainly located at $C_2$ symmetry sites of the ITO bixbyite structure. $H_{\text{CF}}$ is expanded in terms of the $C_2$ symmetry as \cite{KARBOWIAK2010, wybourne1965spectroscopic}:
\begin{equation}
\resizebox{\columnwidth}{!}{$\begin{split}
 H_{CF}= & B^2_0C^2_0 + B^2_2(C^2_{-2}+C^2_2)+ iS^2_2(C^2_{-2}+C^2_2)+ B^4_0C^2_0\\ 
         & + B^4_2(C^4_{-2}+C^4_2) + iS^4_2(C^4_{-2}+C^4_2) + B^4_4(C^4_{-4}+C^4_4) \\ 
         & +iS^4_4(C^4_{-4}+C^4_4)  +B^6_0C^6_0 + B^6_2(C^6_{-2}+C^6_2) \\
         & +iS^6_2(C^4_{-4}+C^4_4)  + B^6_4(C^6_{-4}+C^6_4) + iS^6_4(C^6_{-4}+C^6_4)\\
         & + B^6_6(C^6_{-6}+C^6_6) + iS^6_6(C^6_{-6}+C^6_6)
\end{split}$}
\end{equation}
The total number of CF parameters is 15, considering the real and imaginary parts. However, by choosing an appropriate axis rotation, the parameter $S^2_2$ can be fixed to zero, the $z$-axis is the most common choice \cite{BOSCO2019100028,ANTICFIDANCEV200282, KARBOWIAK2010}.
\begin{table}[b!]
         \caption{Crystal field energies (in $\unit{\per\centi\metre}$) of the \mbox{Tb$^{3+}$} ion in the indium tin oxide (ITO) determined based on the emission spectra from the $^5\text{D}_4$ multiplets taken at $\qty{87}{\kelvin}$. For $f-f$ transitions in sites of $C_2$ local symmetry $2J+1$ lines are expected.}
      \begin{ruledtabular}
     \begin{tabular}{cccccc}
    SLJ Level & Exp & Fit & SLJ Level&  Exp & Fit \\
    \hline
    ${^7\text{F}_6}$  & 0   & -6.6 &                    &3398 & 3402 \\ 
                      & -   & 18.6 &                    &3522 & 3528 \\
                      & -   & 42.3 &                    &3584 & 3585 \\
                      & 107 & 109  &                    & -   & 3604 \\   
                      & -   & 110  &                    & -   &3686 \\
                      & 147 & 148  &                    &3783 &3725 \\
                      & -   & 150  &                    &3852 &3850 \\
                      & 362 & 371  &                    &3906 &3899 \\
                      & -   & 374  & ${^7\text{F}_3}$   &4023 & 4028 \\                   
                      & -   & 511  &                    &  4080 & 4076\\
                      & -   & 512  &                    & 4331 & 4329 \\
                      & -   & 646  &                    & 4347 & 4333 \\
                      & 659 & 655  &                    &4532& 4534 \\
    ${^7\text{F}_5}$  & -   & 717  &                    & - & 4629 \\
                      &2138 & 2154 &                    & 4667& 4660\\
                      &2194 & 2162 & ${^7\text{F}_2}$   &4906 &4910 \\                   
                      &2302 & 2311 &                    &5053& 5039\\
                      &2373 & 2384 &                    & 5078&5072 \\
                      & -   & 2509 &                    &5202 & 5222 \\
                      & -   & 2644 &                    &5287& 5284\\
                      &2795 & 2794 & ${^7\text{F}_1}$   & - & 5424 \\ 
                      &2875 & 2868 &                    & - & 5457 \\
                      &3026 & 3027 &                    & - & 5650 \\
                      &  -  & 3142 &  ${^7\text{F}_0}$  &-  & 5656 \\                  
    ${^7\text{F}_4}$  &3327 & 3332 &                    &   & \\                          
      \end{tabular}
    \end{ruledtabular}
     \label{Energylevels}
 \end{table}
 
The starting values for the free-ion parameters were taken from Carnall et al. \cite{Carnall89} for \mbox{LaF$_3$:Tb$^{3+}$}, $F^2=\qty{88995}{\per\centi\metre}$ , $F^4=\qty{62919}{\per\centi\metre}$, $F^6=\qty{47252}{\per\centi\metre}$, and $\zeta_{4f}=\qty{1707}{\per\centi\metre}$, as well as the correction and correlation factors $\alpha, \beta, \gamma, T^i, M^h, P^k$. The latter were kept fixed throughout the entire fitting procedure since this parameters only change approximately 1\% for different lattices \cite{Liu2005}. The initial values of the CF parameters were taken from Leavitt et al. \cite{Leavitt82}: $B^2_0=\qty{-186}{\per\cm}$, $B^2_2=\qty{-695}{\per\cm}$, $B^4_0=\qty{-1329}{\per\cm}$, $B^4_2=\qty{-1286}{\per\cm}$, $S^4_2=\qty{192}{\per\cm}$, $B^4_4=\qty{727}{\per\cm}$, $S^4_4=\qty{-921}{\per\cm}$, $B^6_0=\qty{258}{\per\cm}$, $B^6_2=\qty{306}{\per\cm}$, $S^6_2=\qty{68}{\per\cm}$, $B^6_4=\qty{459}{\per\cm}$, $S^6_4=\qty{-348}{\per\cm}$, $B^6_6=\qty{-33}{\per\cm}$, and $S^6_6=\qty{-35}{\per\cm}$.
These values correspond to the CF parameters of \mbox{Tb$^{3+}$} at $C_2$ symmetry sites of Y$_2$O$_3$, with a root mean square deviation (rms) of $\sigma = \qty{7.0}{\per\centi\metre}$. Initially, the free-ion parameters were kept fixed while the CF parameters were allowed to vary. In the subsequent iteration, the fitting process was performed using the calculated CF parameters  as starting values, and allowing the free-ion parameters to vary. The experimental and the calculated energy levels are listed in Table \ref{Energylevels}, and the optimized set of the fitted free-ion and CF parameters are given in Table \ref{fitparameters}. CF calculations show good overall agreement between the calculated and experimental CF energy levels, resulting in an rms deviation of  $\sigma = \qty{15.1}{\per\centi\metre}$.

The obtained CF parameters are similar to those found for \mbox{Tb$^{3+}$} in Y$_2$O$_3$ \cite{Leavitt82}, \mbox{Eu$^{3+}$} in Eu$_2$O$_3$\cite{BOSCO2019100028}, and \mbox{Eu$^{3+}$} in In$_2$O$_3$ \cite{ANTICFIDANCEV200282} (\mbox{Eu$^{3+}$} has also 49 CF energy levels related to the $^7F_J$ multiplet). In all these crystals the rare earth ion is in $C_2$ symmetry sites in a bixbyite structure. The parameters are also similar to those for \mbox{Tb$^{3+}$} at $C_2$ symmetry sites in ortho-aluminate TbAlO$_3$ with a perovskite structure  \cite{RUDOWICZ2009, GRUBER2008}.

\begin{table}[t!]
    \centering
    \caption{Free Ion and crystal field parameters (in $\unit{\per\centi\metre}$) for {\mbox{Tb$^{3+}$}} in the $C_2$ site of $Ia\bar{3}$ ITO thin film at $\qty{83}{\kelvin}$. $\sigma$ is the root mean square (rms) deviation between calculated and observed energies. The uncertainty associated with each parameter is indicated in parenthesis. Parameters in brackets were kept fixed during the fitting. ($M^2=0.56M^0$, $M^4=0.31M^0$, $P^4=0.5P^2$, and $P^6=0.1P^2$).}
    \begin{ruledtabular}
    \begin{tabular}{cclccl}
     Parameter & \hspace{0.1cm} & Value & Parameter & \hspace{0.1cm} & Value \\
     \hline
      $E_{\text{avg}}$   &  & 67044 (196)       & $B^2_0$   &  & -750 (90) \\
      $F^2$     &  & 95528 (503)            & $B^2_2$   &  & -1006 (63)\\
      $F^4$     &  & 44544 (889)            & $B^4_0$   &  & -2275 (196)\\
      $F^6$     &  & 52352 (555)            & $B^4_2$   &  & -2226 (150)\\
      $\zeta_{4f}$ &  & 902 (19)            & $S^4_2$   &  & 2298 (114)\\
      $T^2$     &  & [320]                      & $B^4_4$   &  & 2163 (225)\\
      $T^3$     &  & [40]                       & $S^4_4$   &  & 1849 (212)\\
      $T^4$     &  & [50]                       & $B^6_0$   &  & -781 (152)\\
      $T^6$     &  & [-395]                     & $B^6_2$   &  & 1899 (214)\\
      $T^7$     &  & [303]                      & $S^6_2$   &  & -4564 (152)\\
      $T^8$     &  & [317]                      & $B^6_4$   &  & 508 (226)\\
      $\alpha$   &  &[18.4]                     & $S^6_4$   &  & -2486 (112)\\
      $\beta$    &  & [-591]                    & $B^6_6$   &  & -2834 (114)\\
      $\gamma$   &  &[1650]                     & $S^6_6$   &  & 276 (411)\\
      $M^0$      &  & [2.39]                    & & & \\
      $P^2$      &  & [373]                     & $\sigma$ & & 15.1 \\   
    \end{tabular}
    \end{ruledtabular}
      \label{fitparameters}
\end{table}  
\section{Summary}
In this work we have analyzed the local symmetry of the trivalent terbium ions embedded in indium tin oxide matrix with bixbyite structure. In this matrix, \mbox{Tb$^{3+}$} could in principle substitute \mbox{In$^{3+}$} ions in two different cationic positions, $d$ or $b$ sites, inducing stress in the lattice. DFT calculations showed that Tb atoms prefer the $C_2$ symmetry at $d$ sites over $S_6$ symmetry at $b$ sites, thus it is highly likely that \mbox{Tb$^{3+}$} takes a low symmetry site in the bixbyite structure i.e. $C_2$ symmetry sites. This symmetry relaxes selection rules allowing electric dipole transitions that are forbidden in the free rare earth ion. 
It was found that the \mbox{$^5\text{D}_4\rightarrow\leftindex^7{\text{F}}_2$} transition exhibits the highest intensity in photoluminescence spectra of the \mbox{Tb$^{3+}$}, providing a unusual red color to the Tb-related light emission spectrum, and can be resolve in five peaks as a result of Stark splitting induced by inhomogeneities in the crystal field surrounding. A total of 30 intra-4$f$ transitions were identified from the Tb-doped ITO photoluminescence emission spectrum under UV excitation (\qty{325}{\nm}) at \qty{83}{\kelvin}. The energy levels corresponding to the $^7\text{F}_J$ ground multiplet were used to compute the crystal field parameters yielding a final standard deviation of \mbox{$\sigma=\qty{15.1}{\centi\per\metre}$}. The values of the crystal field parameters are similar to those found for \mbox{Tb$^{3+}$} at $C_2$ sites in other hosts materials. Thus confirming that light emitting \mbox{Tb$^{3+}$} ions occupy mainly $C_2$ symmetry sites in the ITO bixbyite structure. These results highlight the capability of tunning the color emission of the \mbox{Tb$^{3+}$} ions by controlling their atomic environment.

\section{Acknowledgments}
This work was funded by the Office of Naval Research (ONR), Grant No. N62909-21-1-2034. E. Serquen acknowledges the Peruvian National Council for Science, Technology and Technological Innovation (CONCYTEC) and the Peruvian National Program of Scientific Research and Advanced Studies (PROCIENCIA), funding scheme E073-2023-01, Grant No. PE501085403-2023. K. Lizárraga and P. Venezuela acknowledge thank the CNPq for the grant 153707/2024-0 under the project 406447/2022-5 Materials Informatics. L. R. Tessler and J. A. Guerra recognize to the DARI-PUCP support for travel expenses. The authors are indebted to the Center for Characterization of Materials (CAM-PUCP) facilities where the experimental characterization was conducted.

\bibliography{references}

\end{document}